\documentclass[runningheads,orivec,a4paper]{llncs}
\usepackage{amssymb}
\usepackage{dsfont}
\usepackage{graphicx}
\usepackage{mathtools}
\usepackage{physics}
\usepackage{textcomp}
\usepackage{amsmath}
\usepackage{amsopn}
\usepackage{tabularx,booktabs,subcaption}
\usepackage{hyperref}
\usepackage{array}
\usepackage{upgreek}
\usepackage{multirow}
\usepackage{algorithm}
\usepackage[noend]{algpseudocode}
\newcolumntype{Y}{>{\centering\arraybackslash}X}
\newcolumntype{C}[1]{>{\centering\let\newline\\\arraybackslash\hspace{0pt}}m{#1}}

\usepackage{amsmath,amsfonts,bm}

\DeclareMathOperator*{\cL}{\mathcal{L}}

\def\1{\bm{1}}

\def\rvx{{\mathbf{x}}}
\def\rvy{{\mathbf{y}}}

\def\vmu{{\bm{\mu}}}
\def\vtheta{{\bm{\theta}}}

\def\vq{{\bm{q}}}

\def\vx{{\bm{x}}}
\def\vy{{\bm{y}}}

\DeclareMathAlphabet{\mathsfit}{\encodingdefault}{\sfdefault}{m}{sl}
\SetMathAlphabet{\mathsfit}{bold}{\encodingdefault}{\sfdefault}{bx}{n}

\DeclareMathOperator*{\argmin}{arg\,min}

\begin{document}

\mainmatter
\title{Generalized Wasserstein Dice Score, Distributionally Robust Deep Learning,\\ and Ranger for brain tumor segmentation:\\ BraTS 2020 challenge}
\titlerunning{GWDL, DRO, and Ranger: BraTS 2020 challenge}
%
\author{Lucas Fidon\inst{1}
	\and S\'ebastien Ourselin\inst{1}
	\and Tom Vercauteren\inst{1}}
\authorrunning{Lucas Fidon et al.}
%
\institute{School of Biomedical Engineering \& Imaging Sciences, King's College London, UK}

\maketitle
\begin{abstract}
    Training a deep neural network is an optimization problem with four main ingredients:
    the design of the deep neural network, the per-sample loss function, the population loss function, and the optimizer.
    However, methods developed to compete in recent BraTS challenges tend to focus only on the design of deep neural network architectures, while paying less attention to the three other aspects.
    In this paper, we experimented with adopting the opposite approach.
    We stuck to a generic and state-of-the-art 3D U-Net architecture and experimented with a non-standard per-sample loss function, the generalized Wasserstein Dice loss, a non-standard population loss function, corresponding to distributionally robust optimization, and a non-standard optimizer, Ranger.
    %
    Those variations were selected specifically for the problem of multi-class brain tumor segmentation.
    The generalized Wasserstein Dice loss is a per-sample loss function that allows taking advantage of the hierarchical structure of the tumor regions labeled in BraTS.
    Distributionally robust optimization is a generalization of empirical risk minimization that accounts for the presence of underrepresented subdomains in the training dataset.
    Ranger is a generalization of the widely used Adam optimizer that is more stable with small batch size and noisy labels.
    We found that each of those variations of the optimization of deep neural networks for brain tumor segmentation leads to improvements in terms of Dice scores and Hausdorff distances.
    With an ensemble of three deep neural networks trained with various optimization procedures, we achieved promising results on the validation dataset and the testing dataset of the BraTS 2020 challenge.
    %
    Our ensemble ranked fourth out of $78$ for the segmentation task of the BraTS 2020 challenge
    with mean Dice scores of $88.9$, $84.1$, and $81.4$, and mean Hausdorff distances at $95\%$ of $6.4$, $19.4$, and $15.8$ for the whole tumor, the tumor core, and the enhancing tumor.
    
\end{abstract}

\section{Introduction}

\begin{figure}[th!]
    \centering
    \includegraphics[width=\linewidth]{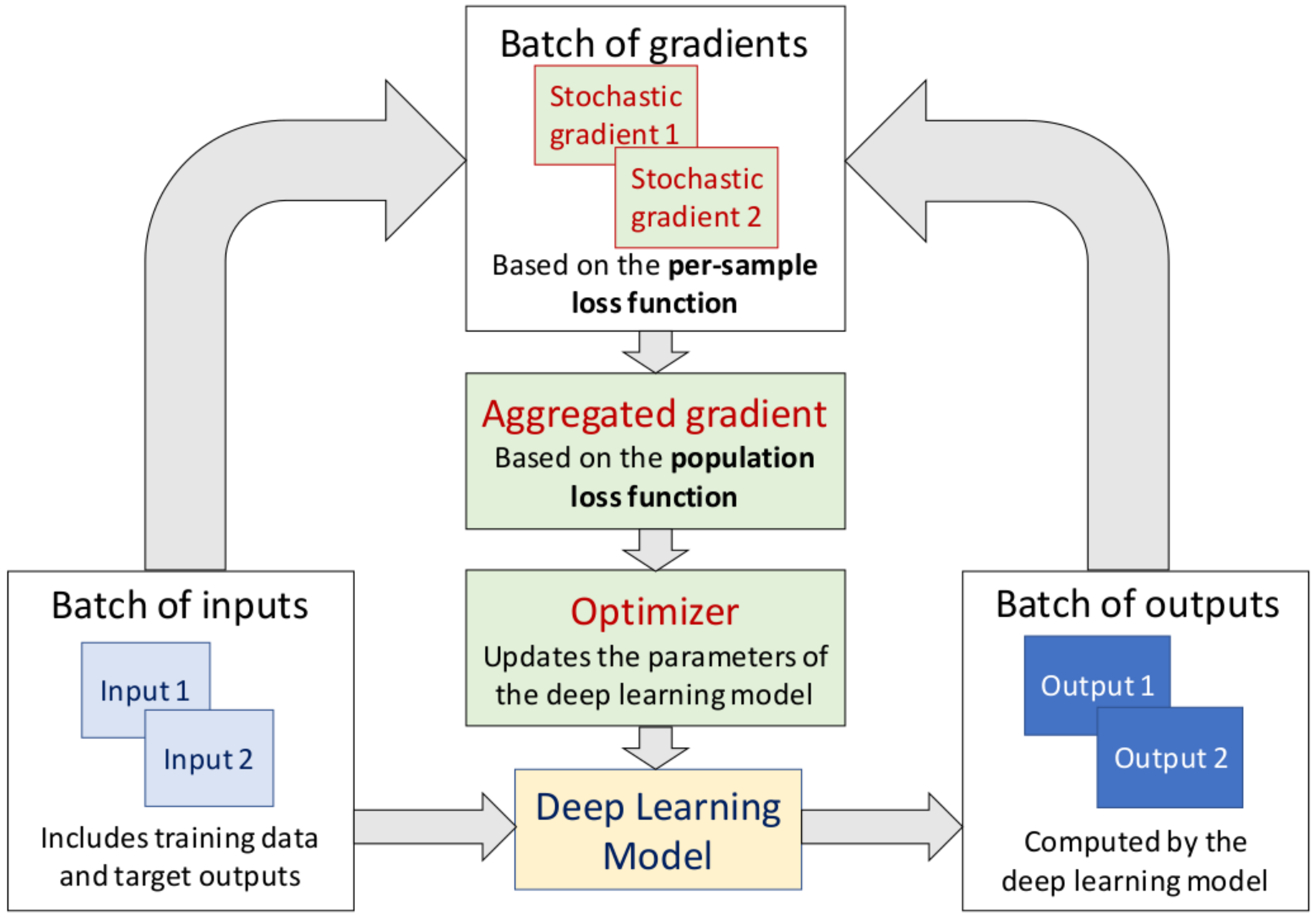}
    \caption{
    \label{fig:diagram_deep_pipeline}
    \textbf{Diagram of a deep learning optimization pipeline.}
    Deep learning optimization methods are made of four main components:
    1) the design of the deep neural network architecture,
    2) the \textbf{per-sample loss function} (e.g. the Dice loss) that determines the stochastic gradient,
    3) the \textbf{population loss function} (e.g. the empirical risk) that determines how to merge the stochastic gradients into one aggregated gradient,
    4) the \textbf{optimizer} (e.g. Adam) that determines how the aggregated gradient is used to update the parameters of the deep neural network at each training iteration.
    In this work, we explore variants for the per-sample loss function, the population loss function and the optimizer for application to automatic brain tumor segmentation.
    }
\end{figure}

Accurate brain tumor segmentation based on MRI is important for diagnosis, surgery planning, follow-up, and radiation therapy~\cite{andres2019po,andres2020dosimetry}. However, manual segmentation is time-consuming (1h per subject for a trained radiologist~\cite{menze2014multimodal}) and suffers from large inter- and intra-rater variability~\cite{menze2014multimodal}.
Automatic and accurate brain tumor segmentation is thus necessary.

In recent BraTS challenges~\cite{bakas2017advancing,menze2014multimodal}, innovations on convolutional neural networks (CNNs) architectures, have led to significant improvement in brain tumor segmentation accuracy~\cite{bakas2018identifying,chandra2018context,fidon2017scalable,niftynet,kamnitsas2016deepmedic,wang2017automatic}.
Recently, the development of nnUNet~\cite{isensee2020automated} has shown that a well-tuned 2D U-Net~\cite{Ronneberger2015} or 3D U-Net~\cite{cciccek20163d} can achieve state-of-the-art results for a large set of medical image segmentation problems and datasets, including BraTS.
The 2D U-Net and 3D U-Net were among the first convolutional neural network architectures proposed for medical image segmentation.
This suggests that the improvement that the design of the deep neural network can bring to brain tumor segmentation is more limited than what was previously thought.
 
In contrast, little attention has been paid to the design of deep learning optimization methods in deep learning-based pipelines for brain tumor segmentation.
We identify three main ingredients other than the design of the deep neural network architecture, in the design of deep learning optimization methods that are illustrated in fig~\ref{fig:diagram_deep_pipeline}:
1) the per-sample loss function or simply \emph{loss function} for short (e.g. the Dice loss~\cite{milletari2016v,sudre2017generalised}),
2) the population loss function (e.g. the empirical risk) whose minimization is hereby referred as the \emph{optimization problem}.
3) the optimizer (e.g. SGD and Adam~\cite{kingma2014adam}),
Recent state-of-the-art deep learning pipelines for brain tumor segmentation uses generic choices of those optimization ingredients such as the sum of the Dice loss and the Cross-entropy loss, Stochastic Gradient Descent (SGD), or Adam as an optimizer and empirical risk minimization.

In this paper, we build upon the 3D U-Net~\cite{cciccek20163d} architecture-based pipeline of nnUNet~\cite{isensee2020automated} and explore alternative loss functions, optimizers, and optimization problems that are specifically designed for the problem of brain tumor segmentation.
We propose to use the generalized Wasserstein Dice loss~\cite{fidon2017generalised} as an alternative per-sample loss function, as discussed in Section \ref{sec:gwdl}, we use distributionally robust optimization~\cite{fidon2020sgd} as an alternative to empirical risk minimization, as discussed in Section \ref{sec:dro}, and we use the Ranger optimizer~\cite{liu2019variance,zhang2019lookahead} as an alternative optimizer, as discussed in Section \ref{sec:ranger}.

The generalized Wasserstein Dice loss~\cite{fidon2017generalised} is a per-sample loss function that was designed specifically for the problem of multi-class brain tumor segmentation. It allows us to take advantage of the hierarchical structure of the tumor regions labeled in BraTS.
In contrast to empirical risk minimization, distributionally robust optimization~\cite{fidon2020sgd} accounts for the presence of underrepresented subdomains in the training dataset.
In addition, distributionally robust optimization does \textbf{not} require labels about the subdomains in the training dataset, such as the data acquisition centers where the MRI was performed, or whether the patient has high-grade or low-grade gliomas.
This makes distributionally robust optimization easy to apply to the BraTS 2020 dataset in which that information is not available to the participants.
Ranger~\cite{liu2019variance,zhang2019lookahead} is a generalization of the widely used Adam optimizer that is more stable with the small batch sizes and noisy labels encountered in BraTS.

Empirical evaluation of those alternatives on the BraTS 2020 validation dataset suggests that they outperform and are more robust than nnUNet.
In addition, our three networks, each one trained with one of the alternative ingredients listed above, appear to be complementary over the three regions of interest in the BraTS challenge: whole tumor, tumor core, and enhancing tumor.
The ensemble formed by our three networks outperforms all of the individual networks for all regions of interest and shows promising results compared to our competitors in the BraTS 2020 challenge.
\textbf{Our ensemble ranked fourth out of $78$ at the segmentation task of the BraTS 2020 challenge} after evaluation on the withheld BraTS 2020 testing dataset.

\section{Method: Varying the Three Main Ingredients of the Optimization of Deep Neural Networks}

In current state-of-the-art deep learning pipelines for brain tumor segmentation, the \textit{training} of the deep neural network consists in the following optimization problem
\begin{equation}
    \label{eq:erm}
    \vtheta^*_{ERM}
    := \argmin_{\vtheta} \frac{1}{n} \sum_{i=1}^n \cL \left(h(\vx_i;\vtheta), \vy_i\right)
\end{equation}
where $h$ is a deep neural network with parameters $\vtheta$, $\cL$ is a smooth per-volume loss function, and $\left\{(\vx_i, \vy_i)\right\}_{i=1}^n$ is the training dataset. 
$\vx_i$ are the input 3D brain MRI T1, T1-gad, T2, and FLAIR volumes, and $\vy_i$ are the ground-truth manual segmentations.

Some of the main ingredients of this optimization problem are:
1) the deep neural network architecture for $h$,
2) the loss function $\cL$,
3) the optimization problem (here \textit{empirical risk minimization}, i.e. we minimize the mean of the per-sample loss functions),
and 4) the optimizer which is the algorithm that allows finding an approximation of $\vtheta^*_{ERM}$.
In recent years, most of the research effort has been put in the deep neural network architecture.
In this work, we set the deep neural network architecture to the 3D U-Net~\cite{cciccek20163d} used in nnUNet~\cite{isensee2020automated}, and explore the three other ingredients.

In this section, we present the per-sample loss function, population loss function, and optimizer that we have used to compete in the BraTS 2020 challenge.

\begin{figure}[tp]
    \centering
    \includegraphics[width=\linewidth]{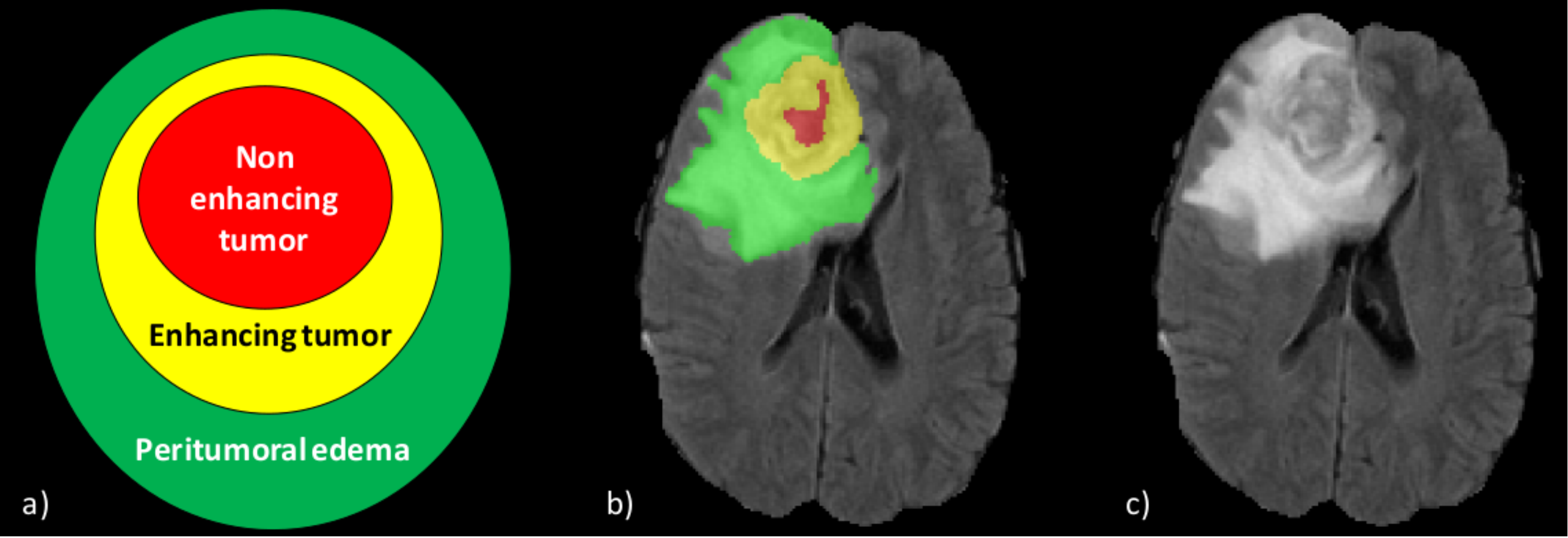}
    \caption{
    \textbf{The brain tumor classes have a hierarchical structure.}
    a) hierarchy of the regions labeled in the BraTS 2020 dataset.
    b) manual segmentation overlaid on the FLAIR image for a case in the BraTS 2020 training dataset.
    c) FLAIR image.
    }
    \label{fig:brats_class_hierarchy}
\end{figure}

\subsection{Changing the Per-sample Loss Function: the Generalized Wasserstein Dice Loss~\cite{fidon2017generalised}}\label{sec:gwdl}
The generalized Wasserstein Dice loss~\cite{fidon2017generalised} is a generalization of the Dice Loss for multi-class segmentation that can take advantage of the hierarchical structure of the set of classes in BraTS.
The brain tumor classes hierarchy is illustrated in fig~\ref{fig:brats_class_hierarchy}.
Our PyTorch implementation of the generalized Wasserstein Dice loss is publicly available\footnote{\url{https://github.com/LucasFidon/GeneralizedWassersteinDiceLoss}}.

When the labeling of a voxel is ambiguous or too difficult for the neural network to predict it correctly, the generalized Wasserstein Dice loss is designed to favor mistakes that are semantically more plausible.
Formally, the generalized Wasserstein Dice loss between the ground-truth (one-hot) class probability map $\textbf{p}$ and the predicted class probability map $\hat{\textbf{p}}$ is defined as~\cite{fidon2017generalised}
\begin{equation}\label{generalised dice}
\mathcal{L}_{GWDL}(\hat{\textbf{p}}, \textbf{p}) = \frac{ 2\sum_{l \neq b} \sum_{i} \textbf{p}_{i,l}(1 - W^M(\hat{\textbf{p}}_i, \textbf{p}_{i}))}{2\sum_{l \neq b}[ \sum_{i} p_{i,l}(1 - W^M(\hat{\textbf{p}}_i, \textbf{p}_{i})) ] + \sum_{i} W^M(\hat{\textbf{p}}_i, \textbf{p}_{i})}
\end{equation}
where $W^M\left(\hat{\textbf{p}}_i, \textbf{p}_i\right)$ is the Wasserstein distance between predicted $\hat{\textbf{p}_i}$ and ground truth $\textbf{p}_i$ discrete probability distribution at voxel i. 
$M= \left(M_{l,l'}\right)_{1 \leq l,\,l' \leq L}$ is a distances matrix between the BraTS 2020 labels, and $b$ is the class number corresponding to the background.

The matrix $M$ informs the generalized Wasserstein Dice loss about the relationships between the classes.
For two classes of indices $l$ and $l'$, the smaller the distance $M_{l,l'}$, the less mistaking a voxel of (ground-truth) class $l$ for the class $l'$ is penalized.

The matrix M is a distance matrix. As a result, it is symmetrical with zeros on its diagonal.
In addition, by convention, we set the maximal-label distance to $1$ that corresponds to the distance between the \textit{background} class and all the other classes.
Specifically, we adapted the distances matrix used in~\cite{fidon2017generalised}, by removing the \textit{necrotic core tumor} that has been merged with the \textit{non-enhancing core} since the BraTS 2017 challenge.
For the classes indices 0:\textit{background}, 1:\textit{enhancing tumor}, 2: \textit{edema}, 3: \textit{non-enhancing tumor}, this corresponds to the matrix
\begin{equation}
    M = 
    \left(
    \begin{array}{cccc}
         0 & 1   & 1   & 1 \\
         1 & 0   & 0.6 & 0.5 \\
         1 & 0.6 & 0   & 0.7 \\
         1 & 0.5 & 0.7 & 0 \\
    \end{array}{}
    \right)
\end{equation}
The distances between the classes reflect the hierarchical structure of the tumor regions, as illustrated in fig~\ref{fig:brats_class_hierarchy}.
The distances between the tumor classes are all lower than $1$ because they have more in common than with the background.

It is worth noting, that since the ground truth segmentation map $\textbf{p}$ is a one-hot segmentation map, for any voxel $i$, we have
\begin{equation}
    W^M\left(\hat{\textbf{p}}_i, \textbf{p}_i\right) = \sum_{l=1}^L p_{i,l}\left(\sum_{l'=1}^L M_{l,l'}\hat{p}_{i,l'}\right)
\end{equation}

\subsubsection*{Previous work:}
Other top performing methods of previous BraTS challenges have proposed to exploit the hierarchical structure of the classes present in BraTS by optimizing directly for the overlapping regions whole tumor, tumor core, and enhancing tumor~\cite{jiang2019two,mckinley2019triplanar,myronenko20183d,wang2017automatic,zhao2019bag}.
However, in contrast to those methods, the generalized Wasserstein Dice loss allows optimizing for both the overlapping regions and the non-overlapping regions labeled in the BraTS dataset simultaneously by considering all the inter-class relationships.

\subsection{Changing the Optimization Problem: Distributionally Robust Optimization~\cite{fidon2020sgd}}\label{sec:dro}

Distributionally Robust Optimization (DRO) is a generalization of Empirical Risk Minimization (ERM) in which the weights of each training sample are also optimized to automatically reweight the samples with higher loss value~\cite{chouzenoux2019general,fidon2020sgd,namkoong2016stochastic,rahimian2019distributionally}.

DRO aims at improving the generalization capability of the neural network by explicitly accounting for uncertainty in the training dataset distribution.
For example, in the BraTS dataset, we don't know if the different data acquisition centers are equally represented.
This can lead the deep neural networks to underperform on the subdomains that are underrepresented in the training dataset.
DRO aims at mitigating this problem by encouraging the neural network to perform more consistently on the entire training dataset.

More formally, DRO is defined by the min-max optimization problem~\cite{fidon2020sgd}
\begin{equation}
    \label{eq:dro}
    \vtheta^*_{DRO}
    := \argmin_{\vtheta}\, \max_{\vq \in \Delta_n}
        \left(
        \sum_{i=1}^n q_i \cL\left(h(\vx_i; \vtheta), \vy_i\right)
        - \frac{1}{\beta} D_{KL}\left(\vq\, \biggr\Vert\, \frac{1}{n}\mathbf{1}\right)
        \right)
\end{equation}
where a new unknown probabilities vector parameter
$\vq
$
is introduced, $\frac{1}{n}\mathbf{1}$ denotes the uniform probability vector $\left(\frac{1}{n}, \ldots, \frac{1}{n}\right)$, $D_{KL}$ is the Kullback-Leibler divergence, $\Delta_n$ is the unit $n$-simplex, and $\beta > 0$ is a hyperparameter.

$D_{KL}\left(\vq\, \biggr\Vert\, \frac{1}{n}\mathbf{1}\right)$ is a regularization term that measures the dissimilarity between $\vq$ and the uniform probability vector $\frac{1}{n}\mathbf{1}$ that corresponds to assign the same weight $\frac{1}{n}$ to each sample like in ERM.
Therefore, this regularization term allows to keep the problem close enough to ERM, and its strength is controlled by $\beta$.

\subsubsection*{Implementation:}
Recently, it has been shown in~\cite{fidon2020sgd} that $\vtheta^*_{DRO}$ can be approximated using any of the optimizers commonly used in deep learning provided the sample volumes are sampled using a \textit{hardness weighted sampling} strategy during training instead of the classic shuffling of the data at each epoch.
For more details on how the hardness weighted probabilities vector $\textbf{q}$ is approximated on-line during training while adding negligible computational overhead, we refer the reader to~\cite[see Algorithm 1]{fidon2020sgd}.

\subsubsection*{DRO and brain tumor segmentation:}
The \textit{hardness weighted sampling} corresponds to a principled hard example mining method and it has been shown to improve the robustness of nnUNet for brain tumor segmentation using the BraTS 2019 dataset~\cite{fidon2020sgd}.
 
In the BraTS dataset, some cases have no enhancing tumor and the Dice score for this class will be either 0 or 1.
As a result, when the mean Dice loss is used as a loss function, those cases with missing enhancing tumor will typically have a higher loss value.
This is an example of cases, perceived as \textit{hard examples} with DRO, that have a higher sampling probability in $\textbf{q}$ during training.

\subsection{Changing the Optimizer: Ranger~\cite{liu2019variance,zhang2019lookahead}}\label{sec:ranger}
Ranger is an optimizer for training deep neural networks that consists of the combination of two recent contributions in the field of deep learning optimization: the Rectified Adam (RAdam)~\cite{liu2019variance} and the lookahead optimizer~\cite{zhang2019lookahead}.
Recently, Ranger has shown promising empirical results for applications in medical image segmentation~\cite{tilborghs2020comparative}.

RAdam~\cite{liu2019variance} is a modification of the Adam optimizer~\cite{kingma2014adam} that aims at reducing the variance of the adaptive learning rate of Adam in the early-stage of training.
For more details, we refer the reader to~\cite[see Algorithm 2]{liu2019variance}.

Lookahead~\cite{zhang2019lookahead} is a generalization of the exponential moving average method that aims at accelerating the convergence of other optimizers for deep neural networks.
Lookahead requires to maintain two sets of values for the weights of the deep neural networks: one set of \textit{fast weights} $\theta$, and one set of \textit{slow weights} $\phi$.
Given a loss function $\mathcal{L}$, an optimizer $A$ (e.g. RAdam), a synchronization period $k$ and a slow weights step size $\alpha > 0$, training a deep neural network with Lookahead is done as follows~\cite[see Algorithm 1]{zhang2019lookahead}
\begin{algorithmic}
\For{$t = 1, 2, \ldots, T $} \Comment{Outer iterations}
  \State $\theta_{t,0} \leftarrow \phi_{t-1}$ \Comment{Synchronize weights}
  \For{$i = 1, 2, \ldots, k$} \Comment{Inner iterations}
    \State $d \sim \mathcal D$ \Comment{Sample a batch of training data}
    \State $\theta_{t,i} \leftarrow \theta_{t,i-1} + A(\mathcal{L}, \theta_{t,i-1}, d)$ \Comment{Update the fast weights}
  \EndFor
  \State $\phi_t \leftarrow \alpha_{t-1} + \alpha \left(\theta_{t,k} - \phi_{t-1}\right)$ \Comment{Update the slow weights}
\EndFor
\Return $\phi_T$
\end{algorithmic}
Lookahead can be seen as a wrapper that can be combined with any deep learning optimizer. However, its combination with RAdam has quickly become the most popular. This is the reason why we considered only lookahead in combination with RAdam in our experiments.

It is worth noting that the optimizers used in deep learning also depend on hyperparameters such as the batch size, the patch size, and the learning rate schedule.
We did not explore in depth those hyperparameters in this work.

\subsection{Deep Neural Networks Ensembling}
Deep neural networks ensembling has been used in previous BraTS challenge to average the predictions of different neural network architectures~\cite{eaton2017ensemble,kamnitsas2017ensembles,lyksborg2015ensemble}.
In this subsection, we discuss the role of ensembling for segmentation using different deep learning optimization methods.

Different deep learning optimization methods can give similarly good segmentations, but they are likely to have different biases and to make different mistakes.
In this case, the ensembling of diverse models can lead to averaging out the inconsistencies due to the choice of the optimization method and improve the segmentation performance and robustness.

Let $\rvx$ be the random variable corresponding to the input 3D brain MRI T1, T1-gad, T2, and FLAIR volumes, and $\rvy$ be the random variable corresponding to the ground-truth manual segmentations for cases with a brain tumor.
%
After training, a deep neural network trained for segmentation gives an approximation
$P(\rvy | \rvx; \vtheta_{\vmu}, \vmu) \approx P(\vy | \vx)$ of the posterior segmentation distribution, where $\vtheta_{\vmu}$ is the vector of trainable parameters of the network obtained after training, and $\vmu$ are the vector of hyperparameters corresponding to the choice of the deep learning optimization method.
Assuming that $P(\rvy | \rvx; \vtheta_{\vmu}, \vmu)$ is an unbiased estimator of $P(\rvy | \rvx)$, and that a set of trained networks corresponding to hyperparameters $\{\vmu_1, \ldots, \vmu_M\}$ are available, an unbiased \textit{ensembling estimation} of $P(\rvy | \rvx)$ with reduced variance is given by
\begin{equation}
    P(\rvy | \rvx)
    \approx
    \frac{1}{M} \sum_{m=1}^M P(\rvy | \rvx; \vtheta_{\vmu_m}, \vmu_m)
\end{equation}

\section{Experiments and Results}

In this section, we first describe the data and the implementation details, and second, we present the models that we compare and analyze their segmentation performance and robustness.

\subsection{Data and Implementation Details}

\subsubsection{Data}
The BraTS 2020 dataset\footnote{See \url{http://braintumorsegmentation.org/} for more details.} has been used for all our experiments. No additional data has been used.

The dataset contains the same four MRI sequences (T1, T1-gad, T2, and FLAIR) for patients with either high-grade Gliomas~\cite{bakas2017HGG} or low-grade Gliomas~\cite{bakas2017LGG}.
All the cases were manually segmented for peritumoral edema, enhancing tumor, and non-enhancing tumor core using the same labeling protocol~\cite{menze2014multimodal,bakas2018identifying,bakas2017advancing}.
The training dataset contains $369$ cases, the validation dataset contains $125$ cases, and the testing dataset contains $166$ cases.
MRI for training and validation datasets are publicly available, but only the manual segmentations for the training dataset are available.
The evaluation on the validation dataset can be done via the BraTS challenge online evaluation platform\footnote{\url{https://ipp.cbica.upenn.edu/}}.
The evaluation on the testing dataset was performed only once by the organizers $48$ hours after they made the testing dataset available to us.
For each case, the four MRI sequences are available after co-registration to the same anatomical template, interpolation to $1$mm isotropic resolution, and skull stripping~\cite{menze2014multimodal}.

\begin{figure}[tp!]
    \centering
    \includegraphics[width=\linewidth]{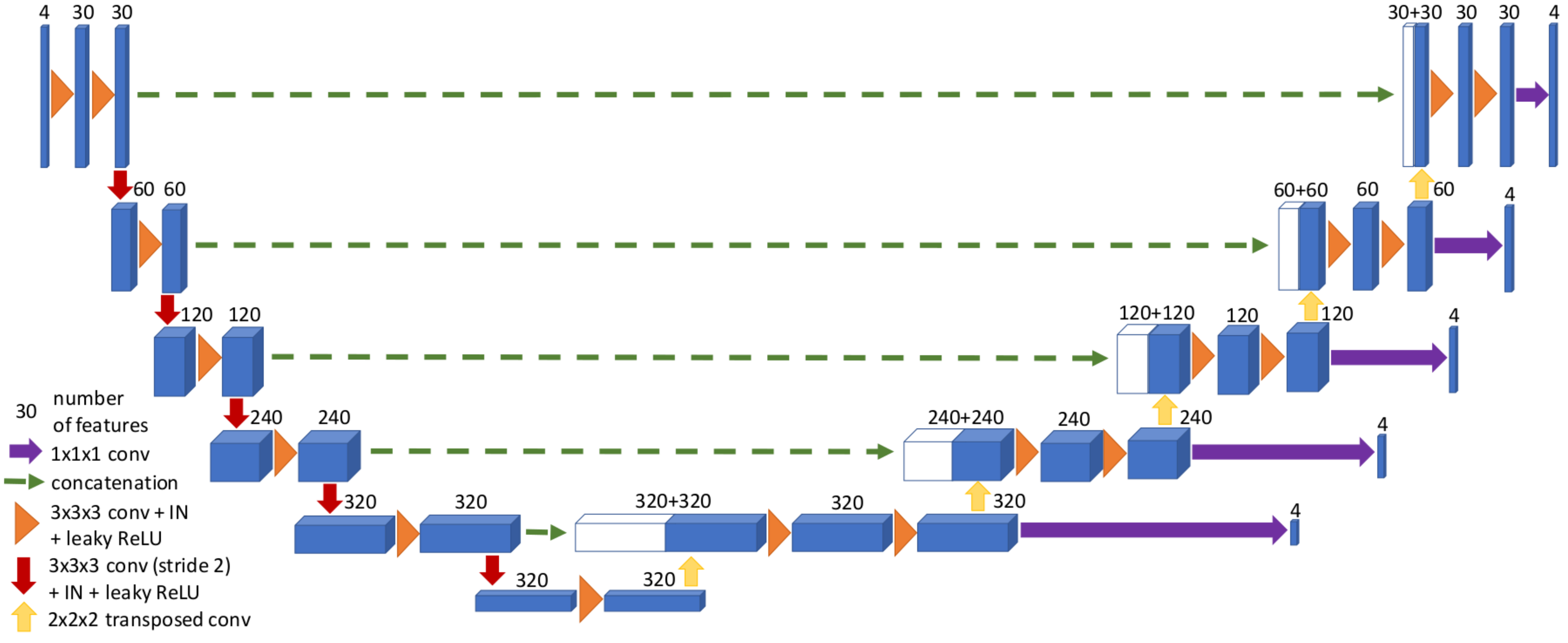}
    \caption{
    \textbf{Illustration of the 3D U-Net~\cite{cciccek20163d} architecture used.}
    Blue boxes represent feature maps. IN stands for instance normalization.
    The design of this 3D U-Net was determined using the heuristics of nnUNet~\cite{isensee2020automated}.
    The main differences between this 3D U-Net and the original 3D U-Net are listed in Section~\ref{sec:unet_details}.
    }
    \label{fig:3dunet}
\end{figure}

\subsubsection{Convolutional Neural Network Architecture}\label{sec:unet_details}
The same 3D U-Net architecture~\cite{cciccek20163d}, based on the heuristics of nnUNet~\cite{isensee2020automated}, was used in all our experiments.
%
%
The main differences compared to the original 3D U-Net~\cite{cciccek20163d} are:
\begin{itemize}
    \item more levels are used,
    \item instance normalization~\cite{ulyanov2016instance} is used instead of batch normalization~\cite{pmlr-v37-ioffe15},
    \item leaky ReLU is used instead of ReLU (with a negative slope of $0.01$),
    \item spatial downsampling is performed using convolutions with stride $2$ instead of average/max-pooling layers,
    \item spatial upsampling is performed using transposed convolutions and the number of features is reduced to match the number of features in the next skip connection before concatenation,
    \item deep supervision~\cite{lee2015deeply} is used (see the purple $1 \times 1 \times 1$ convolutions in fig~\ref{fig:3dunet}),
    \item the maximum number of features is capped at $320$,
    \item the initial number of features is $30$ instead of $32$ (like in nnUNet V1~\cite{isensee2018no}),
    \item the number of features is increased only once in the first level.
\end{itemize}
To help the reader to better appreciate those differences, the 3D U-Net used is illustrated in fig~\ref{fig:3dunet}.

\subsubsection{Training Implementation Details}
Our code is based on the nnUNet code\footnote{\url{https://github.com/MIC-DKFZ/nnUNet} last visited in August 2020.}.
By default and when not indicated otherwise, the sum of the Dice loss and the Cross-entropy loss is used with empirical risk minimization and the SGD with Nesterov momentum optimizer like in~\cite{isensee2020automated}.
The learning rate is decreased at each epoch $t$ as
\[
\lambda_t = \lambda_0 \times \left(1 - \frac{t}{t_{max}}\right)^{0.9}
\]
where $\lambda_0$ is the initial learning rate and $t_{max}$ is the maximum number of epochs fixed as $1000$.
The batch size was set to $2$ and the input patches were of dimension $128 \times 192 \times 128$.
Deep supervision was used as illustrated in fig~\ref{fig:3dunet}.
A large number of data augmentation methods are used: random cropping of a patch, random zoom, gamma intensity augmentation, multiplicative brightness, random rotations, random mirroring along all axes, contrast augmentation, additive Gaussian noise, Gaussian blurring, and simulation of low resolution.
For more implementation details about nnUNet we refer the interested reader to~\cite{isensee2020automated} and the nnUNet GitHub page.

\subsubsection{Inference Implementation Details}
Following nnUNet inference pipeline~\cite{isensee2020automated}, we applied test-time data augmentation, as previously studied in~\cite{wang2019aleatoric}, using flipping along all three spatial dimensions.
When less than $50$ voxels, or equivalently $0.05$mL, in the whole volume were predicted as enhancing tumor, we changed their prediction to non-enhancing tumor.

\subsubsection{Hardware}
GPUs NVIDIA Tesla V100-SXM2 with 16GB of memory were used to train all the deep neural networks.
Training each deep neural network took us between $4$ and $5$ days.

\subsection{Models Description}
In this paragraph, we describe the different models that are compared in Table~\ref{tab:results}.

\subsubsection{nnUNet~\cite{isensee2020automated}}
The original nnUNet code with all the default parameters was trained on the BraTS 2020 training set.
Specifically to the optimization, this means that the sum of the Dice loss and the Cross-entropy loss, SGD with Nesterov momentum, and empirical risk minimization were used for the \textit{nnUNet} model.

\subsubsection{nnUNet + Ranger~\cite{liu2019variance,zhang2019lookahead}}
Exactly the same as for the model \textit{nnUNet} above, except the optimizer was Ranger~\cite{liu2019variance,zhang2019lookahead} with a learning rate of $3 \times 10^{-3}$.

We experimented with different values of the initial learning rate for the Ranger optimizer $\{10^{-3},\, 3 \times 10^{-3},\, 10^{-2}\}$, and the value of $3 \times 10^{-3}$ was retained because it performed best on the BraTS 2020 validation dataset.

We also tried Adam~\cite{kingma2014adam} and RAdam~\cite{liu2019variance} (without lookahead~\cite{zhang2019lookahead}) optimizers, and we tuned the learning rates for each optimizer using the BraTS 2020 validation dataset and the same values for the initial learning rate as mentioned above.
However, we found that Ranger outperformed all the others on the BraTS 2020 validation dataset.

\subsubsection{nnUNet + GWDL~\cite{fidon2017generalised}}
Exactly the same as for the model \textit{nnUNet} above, except the per-sample loss function was the sum of the generalized Wasserstein Dice Loss (GWDL)~\cite{fidon2017generalised} and the Cross-entropy loss.
The initial learning rate was not tuned specifically for use with the GWDL, and we used the default value of nnUNet.

\subsubsection{nnUNet + DRO~\cite{fidon2020sgd}}
Exactly the same as for the model \textit{nnUNet} above, except that we used distributionally robust optimization using the hardness weighted sampler proposed in~\cite{fidon2020sgd}.
The initial learning rate was not tuned specifically for use of DRO and we used the default value of nnUNet.
We choose $\beta=100$ because it is the value that was found to perform best for brain tumor segmentation in~\cite{fidon2020sgd}.

\subsubsection{Ensemble mean softmax}
This model is obtained by averaging the predicted softmax probabilities of the models \textit{nnUNet + Ranger}, \textit{nnUnet + GWDL} and \textit{nnUNet + DRO}.
The model \textit{nnUNet} is not included in the ensemble 
because the model \textit{nnUNet} performed less well than all the other methods in terms of both Dice scores and Hausdorff distances on the three regions of interest.

\begin{table}[tp]
	\centering
	\caption{\textbf{Segmentation results on the BraTS 2020 Validation dataset.}
	The evaluation was performed on the BraTS online evaluation platform.
	The ensemble includes all the single models \textbf{except} the original nnUNet (first model).
	GWDL: Generalized Wasserstein Dice Loss~\cite{fidon2017generalised},
	DRO: Distributionally Robust Optimization~\cite{fidon2020sgd},
	ET: Enhancing Tumor,
	WT: Whole Tumor,
	TC: Tumor Core,
	Std: Standard deviation,
	IQR: Interquartile range.
	Best values restricted to single models are in bold.
	Best values among all models (including ensemble) are in bold and underlined.
	}
	\begin{tabularx}{\textwidth}{c c *{8}{Y}}
		\toprule
        \multicolumn{2}{c}{}
        & \multicolumn{4}{c}{Dice Score ($\%$)} & \multicolumn{4}{c}{Hausdorff $95\%$ (mm)}\\
        \cmidrule(lr){3-6} \cmidrule(lr){7-10}
		\multicolumn{1}{c}{\bf Model} 
		& \multicolumn{1}{c}{\bf ROI} 
		& Mean & Std & Median & IQR
		& Mean & Std & Median & IQR\\ 
	\midrule
	nnUNet~\cite{isensee2020automated}
		& ET 
		& 74.0 & 29.9 & 86.9 & 15.2
		& 38.9 & 109.5 & 2.0 & 2.2\\
		& WT 
		& 90.5 & 7.3 & 92.7 & 6.1 
		& 5.2 & 8.6 & 3.0 & 3.3\\
		& TC
		& 83.9 & 17.0 & 90.1 & 13.7
		& 9.4 & 34.6 & 3.0 & 4.6\\
	\cmidrule(lr){1-10}
		nnUNet & ET 
		 & \textbf{77.4} & 28.2 & 87.6 & 12.1
		 & 32.7 & 101.0 & 1.7 & 2.0\\
		+ & WT 
		 & 90.6 & 7.0 & 92.8 & 6.3
		 & 4.7 & 6.5 & 2.8 & 2.6\\
		Ranger~\cite{liu2019variance,zhang2019lookahead} & TC
		 & 83.8 & 18.1 & 91.3 & 14.3
		 & 9.0 & 34.5 & 2.4 & 4.6\\
	\cmidrule(lr){1-10}
	    nnUNet
		& ET 
		 & 76.7 & 28.0 & 87.4 & 12.6
		 & \textbf{29.8} & 96.1 & 2.0 & 2.0\\
		+ & WT 
		 & \textbf{90.8} & 6.6 & 92.9 & 5.5
		 & \textbf{4.6} & 6.7 & 3.0 & 2.5\\
		GWDL~\cite{fidon2017generalised} & TC
		 & 83.3 & 16.0 & 90.2 & 15.9
		 & 6.9 & 11.4 & 3.2 & 5.3\\
	\cmidrule(lr){1-10}
	    nnUNet & ET 
		 & 75.6 & 28.6 & 87.5 & 12.6
		 & 32.5 & 100.9 & 2.0 & 2.3\\
	   + & WT 
		 & 90.6 & 7.0 & 92.5 & 5.9
		 & \textbf{4.6} & 6.1 & 3.0 & 3.0\\
	   DRO~\cite{fidon2020sgd}	 & TC
		 & \textbf{84.1} & 16.2 & 90.1 & 12.5
		 & \textbf{6.1} & 10.4 & 3.0 & 4.0\\
	\midrule[0.3pt]\midrule[0.3pt]
	    Ensemble & ET 
		 & \underline{\textbf{77.6}} & 27.4 & 87.6 & 11.1
		 & \underline{\textbf{26.8}} & 91.1 & 1.7 & 2.0\\
	    mean & WT 
		 & \underline{\textbf{91.0}} & 6.5 & 92.9 & 6.3
		 & \underline{\textbf{4.4}} & 6.0 & 2.8 & 2.9\\
		softmax & TC
		 & \underline{\textbf{84.4}} & 15.6 & 90.8 & 12.4
		 & \underline{\textbf{5.8}} & 10.2 & 2.8 & 4.3\\
	\bottomrule
	\end{tabularx}
	\label{tab:results}
\end{table}

\begin{table}[tp]
	\centering
	\caption{\textbf{Segmentation results on the BraTS 2020 Testing dataset.}
	The evaluation was performed by the BraTS 2020 organizers.
	ET: Enhancing Tumor,
	WT: Whole Tumor,
	TC: Tumor Core,
	Std: Standard deviation,
	IQR: Interquartile range.
	}
	\begin{tabularx}{\textwidth}{c c *{8}{Y}}
		\toprule
        \multicolumn{2}{c}{}
        & \multicolumn{4}{c}{Dice Score ($\%$)} & \multicolumn{4}{c}{Hausdorff $95\%$ (mm)}\\
        \cmidrule(lr){3-6} \cmidrule(lr){7-10}
		\multicolumn{1}{c}{\bf Model} 
		& \multicolumn{1}{c}{\bf ROI} 
		& Mean & Std & Median & IQR
		& Mean & Std & Median & IQR\\ 
	\midrule
	    Ensemble & ET 
		 & 81.4 & 19.5 & 85.9 & 13.9
		 & 15.8 & 69.5 & 1.4 & 1.2\\
	    mean & WT 
		 & 88.9 & 11.6 & 92.4 & 7.0
		 & 6.4 & 29.0 & 2.9 & 3.0\\
		softmax & TC
		 & 84.1 & 24.5 & 92.6 & 9.2
		 & 19.4 & 74.7 & 2.2 & 2.6\\
	\bottomrule
	\end{tabularx}
	\label{tab:test_results}
\end{table}

\subsection{Mean Segmentation Performance}
Mean Dice scores and Hausdorff distances for the whole tumor, the core tumor, and the enhancing tumor can be found in Table \ref{tab:results}.

In terms of mean Dice scores, 
\textit{nnUNet + DRO} is the only non-ensembling model to outperform \textit{nnUNet} in all regions of interest.
\textit{nnUNet + GWDL} and \textit{nnUNet + Ranger} ouperform \textit{nnUNet} for enhancing tumor and whole tumor.
Among the non-ensembling models, \textit{nnUNet + DRO}, \textit{nnUNet + GWDL} and \textit{nnUNet + Ranger} appear as complementary as they all achieve the top mean Dice score for one of the regions of interest.
That was the motivation for ensembling those three models.

In terms of mean Hausdorff distances,
\textit{nnUNet + DRO}, \textit{nnUNet + GWDL} and \textit{nnUNet + Ranger} outperform \textit{nnUNet} for all regions of interest.

The ensemble outperformed all the other models for all regions in terms of both mean Dice scores and mean Hausdorff distances.

The results of the ensemble on the BraTS 2020 testing dataset are reported in Table~\ref{tab:test_results}. It is those results that were used to rank the different competitors. Our ensemble ranked fourth for the segmentation task.

\subsection{Robustness Performance}
In the summary of the BraTS 2018 challenge, the organizers emphasized the need for more robust automatic brain tumor segmentation algorithms~\cite{bakas2018identifying}.
The authors also suggest using the interquartile range (IQR) of the Dice scores to compare the robustness of the different methods.
IQR for the Dice scores for our models can be found in Table~\ref{tab:results}.

Ensembling and Distributionally Robust Optimization (DRO)~\cite{fidon2020sgd} are two methods that have been empirically shown to decrease the IQR for brain tumor segmentation.
Among the non-ensembling models, \textit{nnUNet + DRO} is the only one to achieve lower Dice scores IQR than \textit{nnUNet} for all the region of interest.
The ensemble achieves the lowest Dice scores IQR for the enhancing tumor and the core tumor regions.

\section{Conclusion}
In this paper, we experimented with three of the main ingredients of deep learning optimization to compete in the BraTS 2020 challenge.

Our results suggest that the segmentation mean performance and robustness of nnUNet~\cite{isensee2020automated} can be improved using distributionally robust optimization~\cite{fidon2020sgd}, the generalized Wasserstein Dice Loss, and the Ranger optimizer~\cite{liu2019variance,zhang2019lookahead}.
Those three features appeared as complementary, and we achieved our top segmentation performance by ensembling three neural networks, each trained using one of them.
In future work, we will explore the combination of those three features to train a single deep neural network.
Our ensemble ranked fourth out of the $78$ participating teams at the segmentation task of the BraTS 2020 challenge after evaluation on the BraTS 2020 testing dataset.

\subsubsection*{Acknowledgments}
This project has received funding from the European Union's Horizon 2020 research and innovation program under the Marie Sk{\l}odowska-Curie grant agreement TRABIT No 765148;
Wellcome [203148/Z/16/Z; WT101957], EPSRC [NS/A000049/1; NS/A000027/1].
Tom Vercauteren is supported by a Medtronic / RAEng Research Chair [RCSRF1819\textbackslash7\textbackslash34].
We would like to thank Luis Carlos Garcias-Peraza-Herrera for helpful discussions and his feedback on a preliminary version of this paper.
We also thank the anonymous reviewers for their suggestions.

%
%
%
\bibliographystyle{splncs04.bst}
\bibliography{main.bib}

\end{document}